\documentclass[journal = jctc, manuscript = letter]{achemso}

\usepackage{amsmath}
\usepackage{amssymb}
\usepackage{graphicx}
\usepackage{mathtools}
\usepackage{braket}
\usepackage{hyperref}
\usepackage{bbold}
\usepackage{calrsfs}
\usepackage{dsfont}
\usepackage{xcolor}
\usepackage{subcaption}

\usepackage{soul}
\let\oldAA\AA
\newcommand{\Angstrom}{\text{\normalfont\oldAA}}

\newcommand{\bz}{\mathbf{z}}
\newcommand{\br}{\mathbf{r}}

\renewcommand{\AA}{\mathbf{A}}

\def\ket#1{| #1 \rangle}
\def\braket#1#2{\langle #1 | #2 \rangle}

\makeatletter
\def\@email#1#2{%
 \endgroup
 \patchcmd{\titleblock@produce}
  {\frontmatter@RRAPformat}
  {\frontmatter@RRAPformat{\produce@RRAP{*#1\href{mailto:#2}{#2}}}\frontmatter@RRAPformat}
  {}{}
}%
\makeatother

\title{Unraveling a cavity induced molecular polarization mechanism from collective vibrational strong coupling}

\author{Dominik Sidler}
  \email{dominik.sidler@psi.ch}
  \affiliation{Laboratory for Materials Simulations, Paul Scherrer Institute, 5232 Villigen PSI, Switzerland }
 \alsoaffiliation{Max Planck Institute for the Structure and Dynamics of Matter and Center for Free-Electron Laser Science, Luruper Chaussee 149, 22761 Hamburg, Germany}
\alsoaffiliation{The Hamburg Center for Ultrafast Imaging, Luruper Chaussee 149, 22761 Hamburg, Germany}%Lines break automatically or can be forced with \\

    \author{Thomas Schnappinger}
  \affiliation{Department of Physics, Stockholm University, AlbaNova University Center, SE-106 91 Stockholm, Sweden}
  
    \author{Anatoly Obzhirov}
  \affiliation{Max Planck Institute for the Structure and Dynamics of Matter and Center for Free-Electron Laser Science, Luruper Chaussee 149, 22761 Hamburg, Germany}
  \alsoaffiliation{The Hamburg Center for Ultrafast Imaging, Luruper Chaussee 149, 22761 Hamburg, Germany}
  
\author{Michael Ruggenthaler}
  \affiliation{Max Planck Institute for the Structure and Dynamics of Matter and Center for Free-Electron Laser Science, Luruper Chaussee 149, 22761 Hamburg, Germany}
  \alsoaffiliation{The Hamburg Center for Ultrafast Imaging, Luruper Chaussee 149, 22761 Hamburg, Germany}

  \author{Markus Kowalewski}
  \email{markus.kowalewski@fysik.su.se}
  \affiliation{Department of Physics, Stockholm University, AlbaNova University Center, SE-106 91 Stockholm, Sweden}

\author{Angel Rubio}
  \email{angel.rubio@mpsd.mpg.de}
  \affiliation{Max Planck Institute for the Structure and Dynamics of Matter and Center for Free-Electron Laser Science, Luruper Chaussee 149, 22761 Hamburg, Germany}
    \alsoaffiliation{The Hamburg Center for Ultrafast Imaging, Luruper Chaussee 149, 22761 Hamburg, Germany}
  \alsoaffiliation{Center for Computational Quantum Physics, Flatiron Institute, 162 5th Avenue, New York, NY 10010, USA}
  \alsoaffiliation{Nano-Bio Spectroscopy Group, University of the Basque Country (UPV/EHU), 20018 San Sebasti\'an, Spain}

\begin{document}

%%%%%%%%%%%%%%%%%%%%%%%%%%%%%%%%%%%%%%%%%%%%%%%%%%%%%%%%%%%%%%%%%%
%                            Abstract                            %
%%%%%%%%%%%%%%%%%%%%%%%%%%%%%%%%%%%%%%%%%%%%%%%%%%%%%%%%%%%%%%%%%%
\begin{abstract}
We demonstrate that collective vibrational strong coupling of molecules in thermal equilibrium can give rise to significant local electronic polarizations in the thermodynamic limit. We do so by first showing that the full non-relativistic Pauli-Fierz problem of an ensemble of strongly-coupled molecules in the dilute-gas limit reduces in the cavity Born-Oppenheimer approximation to a cavity-Hartree equation for the electronic structure. Consequently, each individual molecule experiences a self-consistent coupling to the dipoles of all other molecules, which amount to non-negligible values in the thermodynamic limit (large ensembles). Thus collective vibrational strong coupling can alter individual molecules strongly for localized "hotspots" within the ensemble. Moreover, the discovered cavity-induced polarization pattern possesses a zero net polarization, which resembles a continuous form of a spin glass (or better \textit{polarization glass}). Our findings suggest that the thorough understanding of polaritonic chemistry, requires a self-consistent treatment of dressed electronic structure, which can give rise to numerous, so far overlooked, physical mechanisms.
\end{abstract}

%\date{\today}

\maketitle
 
%%%%%%%%%%%%%%%%%%%%%%%%%%%%%%%%%%%%%%%%%%%%%%%%%%%%%%%%%%%%%%%%%%
%                        Introduction                            %
%%%%%%%%%%%%%%%%%%%%%%%%%%%%%%%%%%%%%%%%%%%%%%%%%%%%%%%%%%%%%%%%%%
%\begin{widetext}

Polaritonic chemistry and materials science is a rapidly growing research field evidenced by a large number of recent review articles.~\cite{ebbesen2016hybrid,ruggenthaler2018quantum,gargiulo2019plasmonic,herrera2020molecular,hirai2020,sidler2022perspective,nagarajan2021chemistry,garcia2021manipulating,nitzan2022polaritons,fregoni2022theoretical,ruggenthaler2022understanding} The strong coupling of matter and light within optical cavities offers a novel way not only to alter and design matter properties, but also to shape the (quantum) properties of light in various ways. For example, magnetic\cite{latini2021ferroelectric} or metal-to-insulator\cite{jarc2022cavity} phase transitions can be altered. Furthermore, cavities can also cause the breakdown of topological protection as reported for the integer quantum Hall effect.\cite{Appugliese2022} In chemistry, the electronic strong coupling, the quantum yield of emissions~\cite{doi:10.1021/jz5004439} or inter-system crossings~\cite{joel2019} can be modified, and photo-chemical reactions can be influenced~\cite{hutchison2012modifying,schwartz2024importance,Felicetti2020-qq,Gudem2021-um,Gudem2022-ej,Couto2022-uv}. For vibrational strong coupling even ground-state (thermally-driven) chemical reactions can be affected~\cite{https://doi.org/10.1002/anie.201605504,hirai2020modulation,hirai2020recent,ahn2023modification}
However, despite a plethora of suggested applications and observed novel effects, we still lack a fundamental understanding of all the relevant underlying microscopic/macroscopic physical mechanisms, specifically in the context of vibrational strong coupling effects.~\cite{sidler2022perspective,doi:10.1021/acs.chemrev.2c00774,campos2022swinging} One of the main reason for this deficiency is the complexity of the full description, which \textit{a priori} requires a holistic approach combining the expertise from different fields of physics and chemistry such as quantum optics, electronic-structure theory, (quantum) statistical mechanics, quantum electrodynamics, molecular and solid state physics.\cite{ruggenthaler2022understanding} Besides questions concerning the observed resonance conditions~\cite{li2021cavity,schafer2021shining,lindoy2023quantum,Rokaj2023}, currently one of the most pressing, unresolved issue in the field is how individual molecules can experience cavity-induced modifications under collective strong coupling.~~\cite{sidler2022perspective,doi:10.1021/acs.chemrev.2c00774,campos2022swinging, ruggenthaler2022understanding} Theoretical attempts to determine how the coupling of the cavity to the ensemble of molecules can modify the chemistry of individual molecules in the thermodynamic limit have so far only been able to describe certain aspects.~\cite{li2020origin,li2020cavity,doi:10.1021/acs.jpclett.2c00974,davidsson2023role} While there have been theoretical suggestions that collective strong coupling can lead to local changes once impurities or (thermally-induced) disorder are introduced in an ensemble~\cite{Schutz2020,sidler2020polaritonic} the existence and nature of such effects for a large ensemble of molecules remained elusive. In this letter we close this important gap by demonstrating numerically that the cavity can indeed induce local polarization effects akin to those observed for small molecular ensembles~\cite{sidler2020polaritonic} for collective coupling in the thermodynamic limit, when treating the many-molecule problem self-consistently within the cavity Born-Oppenheimer approximation of the full Pauli-Fierz theory.

We consider a dilute gas-phase ensemble of $N$ molecules coupled to a photonic environment with confined modes $\alpha$. 
Each molecule consists of $N_e$ electrons and $N_n$ nuclei/ions such that in the long-wavelength limit the Pauli-Fierz Hamiltonian becomes~\cite{spohn2004dynamics,jestadt2019light,tokatly2013time} 
%\begin{widetext}
	\begin{eqnarray}
		\hat{H} &=& \hat{H}_{\rm m} 
		+\frac{1}{2}\sum_{\alpha=1}^{M}\bigg[\hat{p}_\alpha^2+\omega_\alpha^2\Big(\hat{q}_\alpha-\frac{\hat{X}_{\alpha}+\hat{x}_{\alpha}}{\omega_\alpha}\Big)^2\bigg]
		\label{eq:pf_dip_h}
	\end{eqnarray}
%\end{widetext}
with $\hat{H}_{\rm m}$ the usual cavity-free/bare matter Hamiltonian consisting of $N$ molecules. The coupled polarization operators are defined as $\hat{X}_{\alpha}:=\sum_{n=1}^N\sum_{i_n=1}^{N_n} Z_{i_n} \boldsymbol{\lambda}_{\alpha}\cdot\hat{\boldsymbol{R}}_{i_n}$ and $ \hat{x}_{\alpha}:=-\sum_{n=1}^N\sum_{i_n=1}^{N_e}  \boldsymbol{\lambda}_{\alpha}\cdot\hat{\boldsymbol{r}}_{i_n}$, where $Z_{i_n}$ is the nuclear charge and $\hat{\boldsymbol{R}}_{i_n}$ is the coordinate of the $i$-th nucleus/ion of the $n$-th molecule and accordingly for the electrons $\hat{\br}_{i_n}$. The  coupling strength and polarization to the canonical displacement field operators $\hat{q}_\alpha,\hat{p}_\alpha$ with mode frequency $\omega_\alpha$ is defined by $\boldsymbol{\lambda}_\alpha$ and can be obtained from, e.g., macroscopic quantum electrodynamics.~\cite{buhmann2013dispersion,svendsen2023molecules}

In the next step we perform the cavity Born-Oppenheimer approximation (cBOA),\cite{flick2017atoms,flick2017cavity,Flick2018corr} i.e., we treat the electrons of the ensemble as a conditional many-body wave function of all the nuclear degrees of freedom $\underline{\mathbf{R}}$ and all the displacement field coordinates $\underline{q}$.~\cite{latini2021ferroelectric} We subsequently assume the dilute-gas limit and thus the overlaps of \textit{local} many-electron ground-state wave functions $\ket{\chi_n}$ of different molecules is considered negligible, and thus a mean-field ansatz for the \textit{ensemble} electronic wave function
\begin{eqnarray}
	\ket{\Psi} = \ket{\chi_1}\otimes \dots \otimes  \ket{\chi_N}
\end{eqnarray}	    
becomes accurate (see Supporting Information). We note that this ansatz leads to the same set of equations as a Slater determinant of \textit{all} electrons, where we assume that the individual electronic wave functions of different molecules do not overlap. This leads to a set of coupled equations, where the \textit{local} electronic structure of the $n$-th molecule depends self-consistently on all the $N-1$ other molecules. Disregarding bare molecular interaction in the dilute limit, we then have to find the lowest electronic energy state for the following cavity-Hartree equations 

\begin{align}
	\label{eq:CBO_electronic}
 &\left(\hat{H}_{n}(\underline{\mathbf{R}}_n)+\sum_{\alpha=1}^{M}\bigg[\frac{p_\alpha^2}{2}+\frac{\omega_\alpha^2}{2}\Big(q_\alpha-\frac{X_{\alpha}}{\omega_\alpha}\Big)^2 \!\!\!+  \Big(X_{\alpha} - q_\alpha \omega_{\alpha} \!+\!\! \sum_{m\neq n}^{N} \braket{\chi_m}{\hat{x}_{m,\alpha}|\chi_m} \Big) \hat{x}_{n, \alpha}\right. \nonumber \\
 &\qquad +  \left. \frac{\hat{x}_{n,\alpha}^2 }{2}\bigg]\right)\chi_n(\bz_1,...,\bz_{N_e}) = 	\epsilon_n \chi_n(\bz_1,...,\bz_{N_e})
\end{align}	
	 	
for all $N$  molecules simultaneously, i.e., by a self-consistent solution. Eq. \eqref{eq:CBO_electronic} can be solved analytically for an ensemble of $N$ simple harmonic (model) molecules, which is discussed in Ref. \citenum{horak2024analytic}. The bare matter Hamiltonian of a single molecule is defined as $\hat{H}_n$ from $\hat{H}_{\rm m}(\underline{\mathbf{R}})=\sum_{n=1}^N\hat{H}_n(\underline{\mathbf{R}}_n)$ within the dilute-limit approximation and $\hat{x}_{n,\alpha} = -\sum_{i_n=1}^{N_e}  \boldsymbol{\lambda}_{\alpha}\cdot\hat{\boldsymbol{r}}_{i_n}$ is the electronic polarization operator of the $n$-th molecule and $\bz=\br \sigma$ the space-spin variable of an individual electron.  Consequently, the cavity induces  an inter-molecular dipole-dipole energy term  
\begin{align}
	V_{\rm dd} =  \sum_{\alpha=1}^{M}\sum_{n=1}^{N}  \braket{\chi_n}{\hat{x}_{n, \alpha}| \chi_n} \sum_{m \neq n}^{N} \braket{\chi_m}{\hat{x}_{m, \alpha}| \chi_m} \label{eq:ehartree}
 % \sum_{n=1}^{N} \underbrace{\left(\sum_{\alpha=1}^{M} \braket{\chi_n}{\hat{x}_{n, \alpha}| \chi_n}\right)}_{= x_n } \sum_{m \neq n}^{N}\hat{x}_m
\end{align}	
that scales with $N (N-1)$ over the entire ensemble size.  %in addition  to the local quadratic dipole self-energy term. 
This macroscopic scaling is crucial for molecular ensembles, since it counteracts the usual $1/\sqrt{N}$ scaling law of the coupling terms $\boldsymbol{\lambda}_{\alpha}$ for a fixed Rabi splitting, as we show subsequently. 
Indeed, the inter-molecular dipole-dipole interaction $V_{\rm dd}$ is physically straightforward to understand. Because we work in the length gauge, the electric field operator is\cite{schaefer2020relevance} 
\begin{align}
	\hat{\boldsymbol{E}} = 4\pi(\hat{\boldsymbol{D}}-\hat{\boldsymbol{P}}).\label{eq:efield}
\end{align}	 
Notice that the effect of the macroscopic polarization $\hat{\boldsymbol{P}}=-\sum_{\alpha=1}^M\boldsymbol{\lambda}_\alpha(\hat{X}_\alpha+\hat{x}_\alpha)/(4\pi)$ on the microscopic constituents is captured by the dipole self-interaction and scales as $\boldsymbol{\lambda}_{\alpha}^2$.  That these self-interaction terms are important to properly describe the optical response of a material system has been pointed out earlier.~\cite{mukamel1995principles} In addition, it is also established that disregarding this term (as often done in model calculations) and only keeping the cavity-mediated displacement field $\hat{\boldsymbol{D}}=\sum_{\alpha=1}^M\boldsymbol{\lambda}_\alpha \omega_\alpha \hat{q}_\alpha/(4\pi)$ introduces severe theoretical inconsistencies for \textit{ab-initio} simulations.~\cite{rokaj2018light,schaefer2020relevance} In the following,  we will investigate the chemical relevance of treating $\hat{\boldsymbol{P}}$ self-consistently on macroscopic and microscopic scale.

From the self-consistent solution of Eq.~\eqref{eq:CBO_electronic} for all $N$ molecules, we obtain the classical forces for the nuclei/ions and the displacement-field coordinates. In more detail, we can perform an ab-initio molecular dynamics simulation on the polarization-dressed ground-state potential energy surface. To account for classical canonical equilibrium conditions at temperature $T$, which are relevant for many ground-state chemical processes, the classical Langevin equations of motion are propagated, i.e.,~\cite{hutter2012car,sidler2022perspective}
\begin{eqnarray}
M_{i_n}\ddot{\boldsymbol{R}}_{i_n}&=&-\partial_{\boldsymbol{R}_{i_n}} H_n^{\rm n} - \langle\partial_{\boldsymbol{R}_{i_n}} \hat{H}_n^{\rm e}\rangle_0+\sum_{\alpha=1}^M Z_{i_n}\boldsymbol{\lambda}_{\alpha}\Big(\omega_\alpha q_\alpha- X_\alpha - \langle\hat{x}_\alpha \rangle_0\Big)\nonumber\\
&&-\gamma M_{i_n} \dot{\boldsymbol{R}}_{i_n}+\sqrt{2 M_{i_n} \gamma k_B T} \boldsymbol{S}_{i_n}\label{eq:langevin_m}\\
\ddot{ q}_\alpha&=&-\omega_\alpha^2 q_\alpha+\omega_\alpha X_\alpha+\omega_\alpha \langle \hat{x}_\alpha\rangle_0 -\gamma \dot{q}_{\alpha}+\sqrt{2 \gamma k_B T} S_\alpha.
%\langle S(t)\rangle&=&0=\langle S_\alpha(t)\rangle \label{eq:stochmean}\\
%\langle S(t)S(t^\prime)\rangle&=&\delta(t-t^\prime) = \langle S_\alpha(t)S_\alpha(t^\prime)\rangle\label{eq:stochdelta}
\end{eqnarray}
These equations implicitly assume that the Hellmann-Feynman theorem applies, i.e., that the cavity Hartree equations are not only globally but also locally solved using a variational method.  
The bare matter Hamiltonian $\hat{H}_n$ is separated into a classical part, describing nuclear interactions $H_n^{\rm n}$, and the quantized electronic part $\hat{H}_n^{\rm e}$ that parametrically depends on the nuclear positions. Furthermore, we have introduced nuclear masses $M_{i_n}$, friction constant $\gamma$ and component-wise delta-correlated Gaussian noise terms, i.e., $\langle S(t)\rangle=0$, $\langle S(t)S(t^\prime)\rangle=\delta(t-t^\prime)$. Each degree of freedom possesses its individual independent stochastic noise term indexed by $i_n$ and $\alpha$ respectively. We note here that treating the displacement coordinates classically with thermal noise means that we consider photonic excitations due to free charges to be in a classical thermal state~\cite{schaefer2020relevance}.

\begin{figure}
     \begin{subfigure}[b]{0.45\textwidth}
         \centering
         \includegraphics[width=\textwidth]{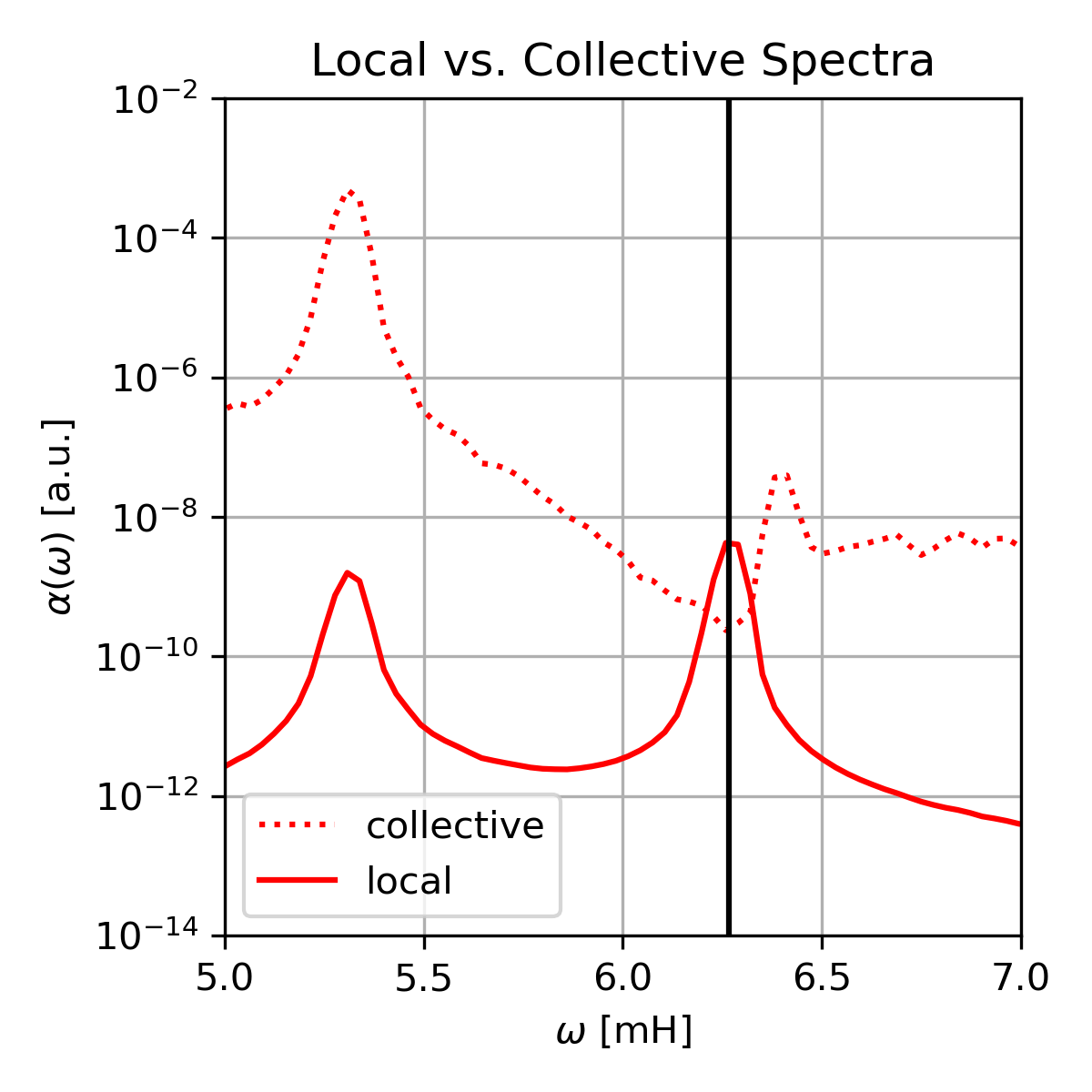}
         \caption{}
         \label{fig:rabisplitting}
     \end{subfigure}
          \begin{subfigure}[b]{0.45\textwidth}
         \centering
         \includegraphics[width=\textwidth]{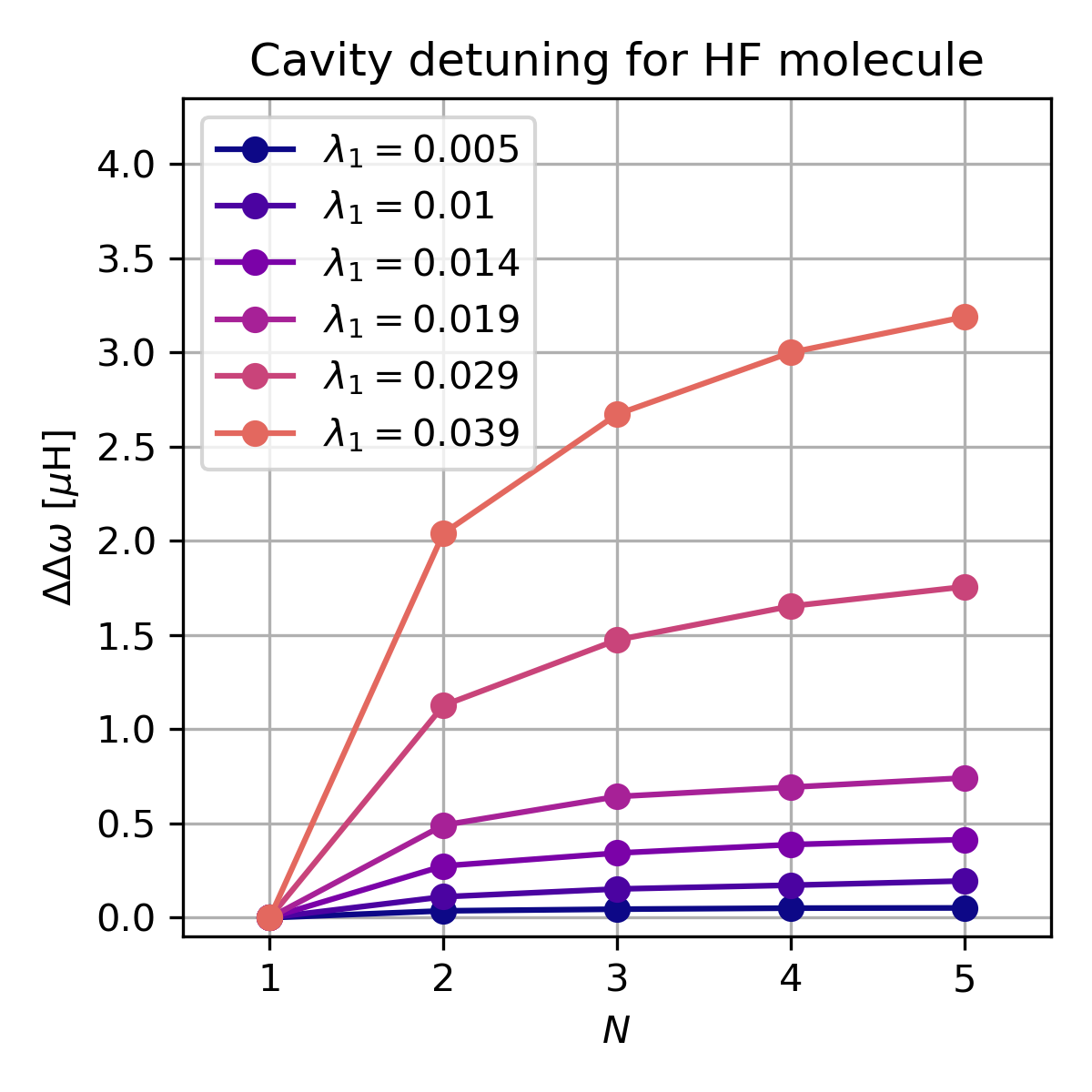}
         \caption{}
         \label{fig:shift}
     \end{subfigure}
        \caption{(a) Vibrational absorption spectra $\alpha(\omega)$ for 900 randomly oriented Shin-Metiu molecules under collective vibrational strong coupling in canonical equilibrium at $k_B T=0.5$ [mH]. The cavity is tuned to the first vibrational mode of the uncoupled molecule at $\omega_\alpha=6.27$ [mH] (black vertical line) with $\lambda_\alpha=0.0085$. The collective Rabi splitting (dotted line) is calculated from the fluctuations of the total ensemble dipole moment (see SI) and shows an asymmetric splitting (red-shifted cavity). In addition, local molecular vibrations (bold line) are monitored in a similar way (see SI), which reveals a significant fraction of individual molecules that are locally strongly coupled, i.e. that vibrate at the frequency of the lower polariton. Furthermore, the local spectrum also indicates that the dark states at $\omega=\omega_\alpha$ are strongly populated. In contrast, no local populations of the upper polaritonic states could be detected at the given temperature. (b) Relative red-shifted cavity frequencies $\Delta\Delta \omega=|\Delta\omega(N)-\Delta\omega(1)|$ with respect to the single molecule shift $\Delta\omega(1)$  for a few perfectly parallelly aligned HF molecules. The collective Rabi splitting was kept constant with respect to $N$ for each chosen $\lambda_1$ by re-scaling $\lambda=\lambda_1/\sqrt{N}$ throughout the computations. The cavity is tuned to the first vibrational mode of the uncoupled HF at $\omega_\alpha=20.35$ [mH] (see SI for further details).  Notice the de-tuning is about two orders of magnitude smaller for the HF molecule  than for the Shin-Metiu molecule. However, the results agree qualitatively, since they suggests a similar finite collective detuning in the large $N$ limit. The overall very small shift is a consequence of the very low polarizability $0.8$ [$\Angstrom^3$] of the HF molecules.\cite{GUSSONI1998163} }
        \label{fig:delta_re}
\end{figure}

\begin{figure}
     %\begin{subfigure}[b]{0.45\textwidth}
         \centering
         \includegraphics[width=0.6\textwidth]{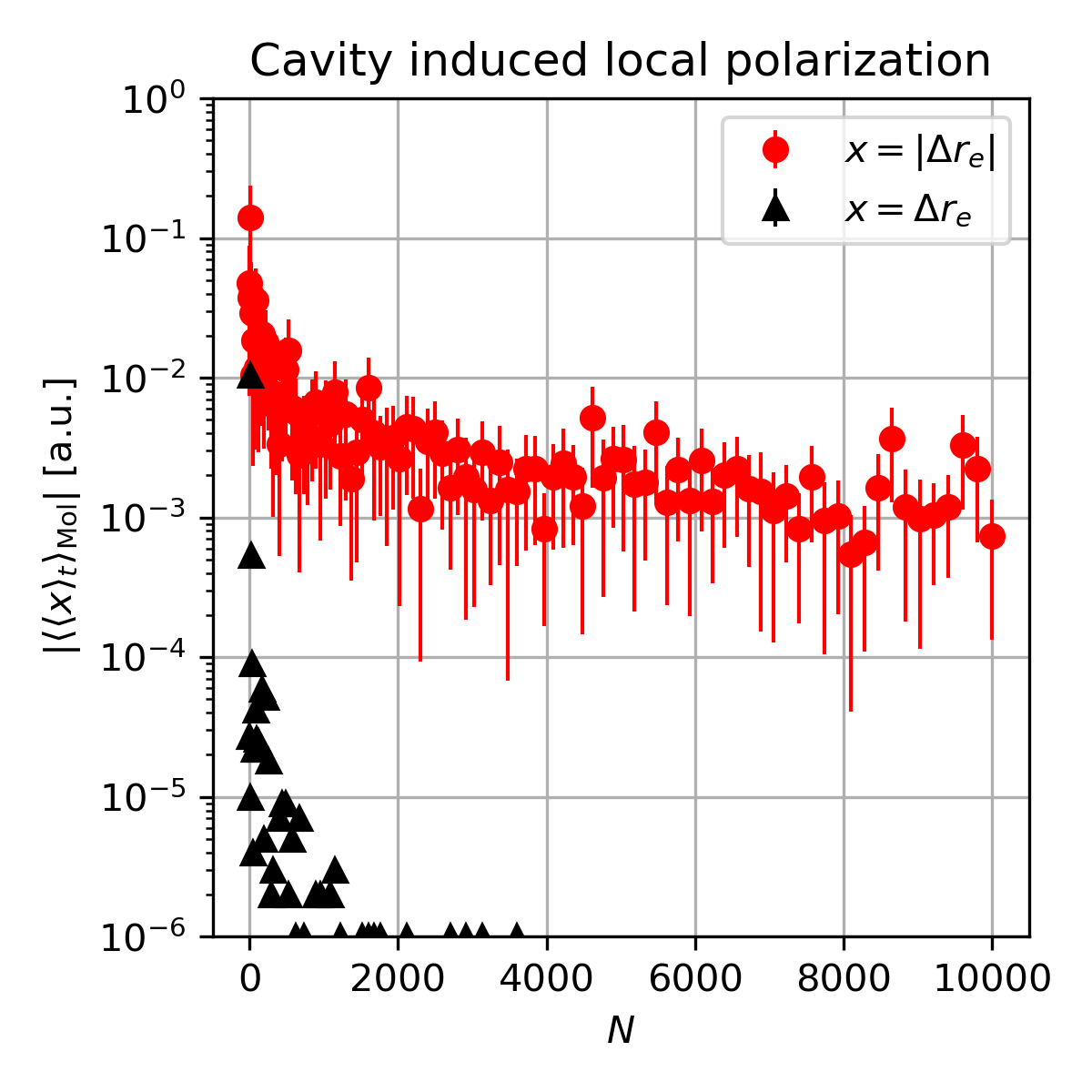}
         \caption{Statistical evaluation of cavity-induced local electronic changes $\Delta r_n(t)=\langle \hat{r}_n\rangle_{\lambda}-\langle \hat{r}_n\rangle_{\lambda=0}$ of the $n$-th molecule with respect to the bare Shin-Metiu molecule in canonical equilibrium at $k_B T=0.5$ [mH] for randomly-oriented molecules (see Supporting Information for details). The Rabi-splitting was kept constant when increasing the number of molecules by choosing a re-scaled $\lambda_\alpha(N)=0.256/\sqrt{N}$. By monitoring $|\Delta r_n|$ (red dots), we observe a non-zero saturation of the cavity induced local polarizations in the large-$N$ limit, where the standard deviations with respect to different molecules are indicated by vertical red lines. At the same time, the total polarization of the ensemble, which is related to $\Delta r_n$ (black triangles), quickly approaches zero, since the cavity cannot induce a non-zero polarization of the ensemble.    
         Consequently, our simulations suggest that cavity-induced local strong coupling effects persist in the thermodynamic limit $(N \gg 1)$ of a molecular ensemble under collective vibrational strong coupling. In other words, the self-consistent treatment of Eq.~\eqref{eq:CBO_electronic} is decisive to describe ground-state polaritonic chemistry accurately for realistic molecular ensembles. }
         \label{fig:dre}
     %\end{subfigure}
%          \begin{subfigure}[b]{0.45\textwidth}
%         \centering
%         \includegraphics[width=\textwidth]{Delta_re_t_mol_N_temperature_00005.png}
%         \caption{}
%         \label{fig:y equals x}
%     \end{subfigure}
\end{figure}

In the first step, the collective Rabi splitting is calculated for a model system consisting of $N=900$ randomly-oriented and slowly rotating Shin-Metiu molecules~\cite{shin1995} strongly coupled to a single cavity mode $\omega_\alpha$ at $T=158$ K, yielding a clear lower and upper polaritonic resonance, as depicted by the dotted line in Fig.~\ref{fig:rabisplitting} (see SI for further details). Notice, the differently aligned molecules in the dilute limit can also be re-interpreted as aligned molecules coupled to a spatially modulated displacement field, which makes our results more generally applicable.   Taking into account the self-consistent treatment of the cavity-induced molecular polarization $\hat{\boldsymbol{P}}$ leads to a detuning of the cavity towards lower frequencies, which manifests itself in the asymmetric splitting with respect to the bare cavity mode $\omega_\alpha$, indicated by the black vertical line. The redshift of the cavity frequency can directly be related to the refractive indesx of the ensemble within the harmonic approximation.\cite{horak2024analytic,fiechter2024understanding} Simulations of a few less polarizable hydrogen fluoride (HF) dimers\cite{schnappinger2023cavitybornoppenheimer} show the same tendency as demonstrated in Fig.~\ref{fig:shift} and Fig.~S3, but of significantly smaller magnitude. Qualitatively similar results can also be found for perfectly aligned Shin-Metiu molecules (see Fig. S1 in the SI). The observed cavity induced de-tuning resembles the dipole-dipole interaction induced Lorentz redshift within dense atomic ensembles (in absence of a cavity), which depends strongly on the microscopic polarizability of the media, similar to our result.\cite{Maki1991,Javanainen2014} Notice that when tuning the cavity to much lower frequencies (e.g. ro-vibrational regime), the presence (back-action) of permanent molecular dipoles is expected to significantly contribute to the red-shift alongside with the molecular polarizability. This dynamic re-orientation contribution is neglected in our simulations. The observed collectively induced redshift of a filled cavity with respect to a bare one has also been seen in experiments~\cite{thomas2016ground} and may in principle be simply approximated by a suitably chosen refractive index of the ensemble~\cite{sidler2022perspective}, with perfect agreement for harmonic molecules as discussed in Ref. \citenum{horak2024analytic}. We also note, that here we get the redshift directly from the simulation, where we calculate the self-consistent polarization of the ensemble of molecules. That is, we calculate implicitly the ensemble polarizability and its back-action on the cavity mode. In the case of free-space modes, this is the standard way to determine the refractive index of a material.~\cite{mukamel1995principles,ullrich2011time,marques2012fundamentals} 
Now the question arises whether the accurate self-consistent and microscopic treatment of the polarization can additionally induce local field effects that cannot be disregarded in the thermodynamic limit ($N\gg1$) and that are not captured by a simple refractive index picture? Earlier evidence for collective electronic strong coupling for a few molecules indeed indicates that there might be such an effect,~\cite{sidler2020polaritonic} yet the existence of similar local polarization effects for a thermal ensemble under vibrational strong coupling in the large-$N$ limit remained unclear.  
As the first local observable, we analyze the local molecular dipole vibrations for individual Shin-Metiu molecules, which reveals a (locally) populated lower polariton (solid line in Fig.~\ref{fig:rabisplitting}) and a strongly populated dark state at $\omega=\omega_\alpha$. A local upper polariton could not be identified, i.e., may be too weakly populated to overcome the thermal broadening for the given system. Simulations show that the usual $\sqrt{N}$-collective Rabi split scaling law of the Tavis-Cumming model remains preserved collectively as well as locally, when including local polarization effects self-consistently (see Fig. S1 of the Supporting Information). As the second local observable, we propagate the system self-consistently and measure at every time step, i.e., for every realized classical configuration $(\underline{R},q_\alpha)$, the difference between the exact solution of Eq.~\eqref{eq:CBO_electronic} and the electronic bare matter problem by monitoring $\Delta r_n(t)=\langle \hat{r}_n\rangle_{\lambda}-\langle \hat{r}_n\rangle_{\lambda=0}$ in the electronic ground-state. This allows to measure cavity-induced local polarization effects in thermal equilibrium, since the full electronic problem reduces to the bare local matter problem in the thermodynamic limit if only the displacement field is considered instead. Our simulation results in Fig.~\ref{fig:dre} reveal a nonvanishing cavity-induced local ensemble polarization, i.e., $\langle |\Delta r_n|\rangle\neq 0$, that persist even in the large-$N$ limit. At the same time, the total electronic polarization of the ensemble remains zero, i.e., $\langle \Delta r_n\rangle= 0$, as expected from the symmetry of Eq.~\eqref{eq:pf_dip_h}.  Consequently, our numerical results show that the chemical properties of individual molecules can be locally modified by collective strong vibrational coupling to the cavity. Fundamentally speaking, our observation of a continuous distribution of cavity induced molecular polarizations with zero net polarization resembles a continuous form of a spin-glass\cite{mezard1987spin} (or better \textit{polarization-glass}). The continuous distribution automatically implies the existence of \textit{hotspots} within the molecular ensemble, where the collective coupling can strongly polarize single molecules and thus significantly alter their chemical properties. Hoever, the average cavity-induced polarization remains rather small, which seems in agreement with recent NMR experiments.\cite{ebbesen2024} Notice this collectively induced local mechanism occurs without external driving, i.e. the sole presence of a thermal bath is sufficient.  Analogous results hold also for perfectly aligned molecules, as shown in Fig. S2 of the Supporting Information. Physically, the appearance of local strong-coupling effects can be understood by interpreting the local polarization in a dipole picture, as previously done for electronic strong coupling of a few nitrogen dimers~\cite{sidler2020polaritonic}. While the total (macroscopic) polarization is zero, non-trivial local dipole modifications are possible for heterogeneous systems that can still cancel each other, i.e., as seen from the sum $m \neq n$ in Eq.~\eqref{eq:CBO_electronic}. This local polarization induces a mirror dipole in the rest of the ensemble. At this point, we would like to highlight the relevance of (random) disorder in the ensemble (temperature and/or different molecular orientations, vibrational states and polarizabilities), which enables a heterogeneous structure of modified local polarizations that can cancel collectively in analogy to a spin glass\cite{sherrington1975}. For atoms, which do not have a static dipole moment, no local effect is expected, as can be confirmed by simulating a small ensemble of up to five neon atoms (see Fig~S4 in the Supporting Information). In other words, having spherically symmetric systems without (different) internal nuclear degrees of freedom, all local dipole contributions will be equivalent and thus the local polarization needs to be zero, in order to have a zero macroscopic polarization. Furthermore, the simple harmonic model considerations in Ref. \citenum{horak2024analytic} demonstrate another important ingredient for the formation of polarization glass, which is the presence of a complex electronic structure, i.e., anharmonic electron interactions (e.g. Coulomb) are mandatory. Overall, our results do not contradict well-established knowledge from quantum-optical models for \textit{atomic} systems. However, they show that for molecular ensembles, the formalism becomes more involved, and the self-consistent (!) treatment of the local polarization may become decisive in capturing all relevant aspects of polaritonic chemistry. We also note that the free-space mode structure of the electromagnetic field, which is homogeneous and isotropic, is not able to test for disorder in the same way as cavity modes do by having preferred polarization directions and frequencies. This breaking of symmetry explains why similar effects are not expected for coupling to free-space modes.  

To conclude, we have reformulated the computationally inaccessible many-molecule Pauli-Fierz problem of polaritonic chemistry in terms of an efficient cavity-Hartree many-molecule problem, within the dilute-gas limit and the cBOA. 
Simulating the corresponding Langevin equations of motion under vibrational strong coupling in thermal equilibrium reveals that  solving the cavity-Hartree equations self-consistently, and thus including dipole-dipole interactions between molecules, can be decisive to capture all relevant aspects of polaritonic chemistry. 
The reason is that non-trivial local (on the individual-molecule level) polarization distributions can arise with zero net polarization, which can persist in the thermodynamic limit and thus may be regarded as a \textit{polarization-glass} phase. The continuous distribution implies the existence of molecular \textit{hotspots}, where chemistry is locally altered significantly by the cavity. Furthermore, our self-consistent accounting for ensemble polarization effects lead to a detuning of the cavity towards lower frequencies, which is in line with experimental evidence\cite{thomas2016ground} and shows that the dipole self-interaction term is a necessary ingredient to capture the basic effect of a changed refractive index. 

% Should we say something about the 1/N problem here, that we overcome to some degree now?

The present result may have far-reaching consequences for the theoretical description of polaritonic chemistry and materials science, since they provide a so far overlooked, yet simple and intuitive physical mechanism that can induce local molecular changes in the thermodynamic limit. This local mechanism may be the missing piece to settle current discrepancies between existing simplified models for a macroscopic ensemble of molecules and experiments. Furthermore, our cavity-Hartree equations are well suited to be included in existing computational methods,~\cite{ruggenthaler2014quantum,luk2017multiscale,flick2017cavity,jestadt2019light,chen2019ehrenfest,chen2019ehrenfestII,sidler2020chemistry,haugland2020coupled} which will enable the efficient exploration of the large chemical space with a multitude of observables. Particularly, large ensemble sizes under self-consistent vibrational strong coupling should become accessible by established ab-initio molecular dynamics codes~\cite{luk2017multiscale,chen2019ehrenfest,chen2019ehrenfestII,Schnappinger23jctc} and potentially with the help of self-consistent embedding schemes~\cite{schaefer2022polaritonic}. Last but not least, the existence of a macroscopically induced microscopic polarization mechanism opens many interesting fundamental physical questions.  For instance, can we efficiently control microscopic (quantum) properties of individual molecules via a thermal macroscopic field, or are the experimentally observed modifications of chemical reactions purely due to change in the statistics? On the more theoretical side, can our results be generalized to the liquid or even solid phase under collective strong coupling conditions? What are the physical properties of the suggested cavity induced \textit{polarization glass} phase and its relation to a spin glass? Can thus the thermodynamics of molecules under VSC be affected akin to frustration\cite{mezard1987spin,parisi2006spin} in a spin glass? All these aspects open many interesting questions that lie at the boundaries between physics and chemistry and need the combination of various different viewpoints and methods~\cite{ruggenthaler2018quantum}.

\begin{acknowledgement}
We thank I-Te Lu, Simone Latini, Abraham Nitzan and Gerrit Groenhof for inspiring discussions and helpful comments.
This work was made possible through the support of the RouTe Project (13N14839), financed by the Federal Ministry of Education and Research (Bundesministerium für Bildung und Forschung (BMBF)) and supported by the European Research Council (ERC-2015-AdG694097), the Cluster of Excellence “CUI: Advanced Imaging of Matter” of the Deutsche Forschungsgemeinschaft (DFG), EXC 2056, project ID 390715994 and the Grupos Consolidados (IT1249-19).
The Flatiron Institute is a division of the Simons Foundation.
M.K. acknowledges funding from the European Research Council (ERC) under the European Union’s Horizon 2020 research and innovation program (grant agreement no. 852286).
\end{acknowledgement}

\begin{suppinfo}

\end{suppinfo}
\bibliography{letter}
\end{document}

%\preprint{AIP/123-QED}

\title{SUPPORTING INFORMATION: Unraveling a cavity induced molecular polarization mechanism from collective vibrational strong coupling}

% Force line breaks with \\
\author{Dominik Sidler}
  \email{dsidler@mpsd.mpg.de}
 \affiliation{Max Planck Institute for the Structure and Dynamics of Matter and Center for Free-Electron Laser Science, Luruper Chaussee 149, 22761 Hamburg, Germany}
\affiliation{The Hamburg Center for Ultrafast Imaging, Luruper Chaussee 149, 22761 Hamburg, Germany}%Lines break automatically or can be forced with \\

  \author{Thomas Schnappinger}
  \email{thomas.schnappinger@fysik.su.se}
  \affiliation{Department of Physics, Stockholm University, AlbaNova University Center, SE-106 91 Stockholm, Sweden}

  \author{Anatoly Obzhirov}
  \email{anatoly.obzhirov@mpsd.mpg.de}
  \affiliation{Max Planck Institute for the Structure and Dynamics of Matter and Center for Free-Electron Laser Science, Luruper Chaussee 149, 22761 Hamburg, Germany}
  \affiliation{The Hamburg Center for Ultrafast Imaging, Luruper Chaussee 149, 22761 Hamburg, Germany}
  
\author{Michael Ruggenthaler}
  \email{michael.ruggenthaler@mpsd.mpg.de}
  \affiliation{Max Planck Institute for the Structure and Dynamics of Matter and Center for Free-Electron Laser Science, Luruper Chaussee 149, 22761 Hamburg, Germany}
  \affiliation{The Hamburg Center for Ultrafast Imaging, Luruper Chaussee 149, 22761 Hamburg, Germany}

  \author{Markus Kowalewski}
  \email{markus.kowalewski@fysik.su.se}
  \affiliation{Department of Physics, Stockholm University, AlbaNova University Center, SE-106 91 Stockholm, Sweden}

\author{Angel Rubio}
  \email{angel.rubio@mpsd.mpg.de}
  \affiliation{Max Planck Institute for the Structure and Dynamics of Matter and Center for Free-Electron Laser Science, Luruper Chaussee 149, 22761 Hamburg, Germany}
    \affiliation{The Hamburg Center for Ultrafast Imaging, Luruper Chaussee 149, 22761 Hamburg, Germany}
  \affiliation{Center for Computational Quantum Physics, Flatiron Institute, 162 5th Avenue, New York, NY 10010, USA}
  \affiliation{Nano-Bio Spectroscopy Group, University of the Basque Country (UPV/EHU), 20018 San Sebasti\'an, Spain}

\date{\today}% It is always \today, today,

\maketitle

\section{Cavity Hartree equations}

In order to derive the cavity Hartree Eq.~(3) in the letter, we start from the fully quantize Pauli-Fierz Hamiltonian  given in Eq.~(1) and apply a Born-Huang expansion to the total wave function following Ref. \citenum{flick2017cavity}. In more detail, we separate the electronic degrees of freedom from the nuclear-photonic degrees of freedom yielding $\Psi_i=\sum_{j=1}^\infty \gamma_{ij}(\bold{R},{q}) \psi_j(\bold{r},\bold{R},{q})$ that solves  $\hat{H}\Psi_i=E_i\Psi_i$. Afterwards, we employ the cavity Born-Oppenheimer approximation in its simplest form, which assumes a classical treatment of the nuclear and photonic degrees of freedom.\cite{flick2017cavity} This approximation accurately describes cavity mediated ground-state chemistry, i.e. when the cavity is tuned on the vibrational degrees of freedom and the electronic excitations are sufficiently separated energetically such that non-adiabatic couplings (e.g. conical intersections) are negligible.\cite{flick2017cavity,sidler2022perspective,ruggenthaler2022understanding,Schnappinger23jctc,Fischer2023-pe} Consequently, the system separates into an electronic subsystem $\hat{H}^{\rm e}$, which parametrically depends on the modes $\underline{q}$ and the nuclear position vector $\underline{\boldsymbol{R}}$, and into a nuclear-photonic part $\hat{H}^{\rm npt}$.
Following the notation of the letter, the resulting electronic Hamiltonian operator becomes
\begin{eqnarray}
 \hat{H}^{\rm e}&:=&\hat{H}_{\rm m}^{\rm e}
 +\sum_{\alpha=1}^M\bigg(\frac{1}{2} \hat{x}_\alpha^2+\hat{x}_\alpha X_\alpha-\omega_\alpha  \hat{x}_\alpha q_\alpha\bigg),
\end{eqnarray}
which  depends parametrically on $\bold{R},q_\alpha$.
The corresponding groundstate Hamiltonian of the coupled nuclear-photon degrees of freedom is given by,
\begin{eqnarray}
H^{\rm npt}&:=&H_{\rm m}^{\rm n}+\sum_{\alpha=1}^M\bigg(\frac{p_\alpha^2}{2}
 +\frac{\omega_\alpha^2}{2}\Big(q_\alpha-\frac{X_\alpha}{\omega_\alpha} \Big)^2+\bra{\psi_0}\hat{H}_e(\underline{\bold{R}},\underline{q})\ket{\psi_0}\bigg).%({R},q_\alpha)\\
\end{eqnarray}

In a next step, we demonstrate that the groundstate many-electron eigenvalue problem imposed by $\hat{H}^{\rm e}$ can be solved exactly in the dilute limit by a simple Hartree product ansatz. To show this, we start from a Hartree-Fock mean-field Ansatz given in Eq.~(2) of the letter to solve for $\varepsilon_0=\min E_{HF}=\min \bra{\psi_{HF}}\hat{H}^{\rm e}\ket{\psi_{HF}}$. %with
%\begin{eqnarray}
%\psi_{HF}:=\frac{1}{\sqrt{N}}\begin{vmatrix}
%    \chi_{1}({r}_1) & \dots  &  \chi_{N}({r}_1) \\
%    \vdots & \vdots  & \vdots \\
%    \chi_{1}({r}_N) &  \dots  & \chi_{N}({r}_N),
%\end{vmatrix}
%\end{eqnarray}
%\revDS{spin?}
The resulting Hartree-Fock energy can then simply be written as,
\begin{eqnarray}
E_{HF}&=& \sum_{n=1}^N \bra{\chi_n}\hat{H}_{n}^{\rm e}+\sum_{\alpha=1}^M\bigg[\frac{1}{2}\hat{x}_{n,\alpha}^2+\hat{x}_{n,\alpha} X_\alpha-\omega_\alpha  \hat{x}_{n,\alpha} q_\alpha\bigg]\ket{\chi_n}\nonumber\\
&&+\frac{1}{2}\sum_{n, m}^N \sum_{\alpha=1}^M\bra{\chi_n}\hat{x}_{n,\alpha} \ket{\chi_n}\bra{\chi_m}\hat{x}_{m,\alpha}\ket{\chi_m}
\cancel{-\frac{1}{2}\sum_{n, m}^N\sum_{\alpha=1}^M  \bra{\chi_n}\hat{x}_{n,\alpha} \ket{\chi_m} \bra{\chi_m}\hat{x}_{m,\alpha}\ket{\chi_n}},\label{eq:hartree_en}
\end{eqnarray}
where the cavity-induced electron-electron exchange cancel in the last terms, except for $n=m$, due to the dilute limit assumption, i.e, the electronic structure of different molecules must not overlap. Notice that for non-overlapping electronic structures the inter-molecular exchange and correlation energies are zero, and if the molecules are far apart also the Coulomb Hartree energy goes to zero. Therefore we have $\hat{H}_{\rm m}^{\rm e}$ simplifies to $\hat{H}_{\rm m}^{\rm e}=\sum_{n=1}^N \hat{H}_n^{\rm e}$ in Eq.~\eqref{eq:hartree_en}. Consequently, our mean-field ansatz becomes exact in the dilute limit. However, the exact solution of the local bare matter problem, i.e., finding accurate eigenfunctions of $\hat{H}_{n}^{\rm e}$, may itself be a highly non-trivial problem that requires to consider the intra-molecular correlations with computationally expensive post Hartree-Fock methods. From Eq. (\ref{eq:hartree_en}) we find the resulting Hartree equations for the $n$-th orbital upon variation of the orbitals $\chi$ as presented in Eq. (3) of the letter. 
%\begin{eqnarray}
%\bigg[\frac{\hat{\bold{p}}_i^2}{2m}+W(\hat{\bold{r}}_i,\bold{R}_i)+\frac{1}{2}\hat{X}_{r,i}^2+\hat{X}_{r,i} %X_R-\omega_\alpha  \hat{X}_{r,i} q_\alpha&&\label{eq:hartree}\\
%+e^2\sum_{j\neq i}^N \int \chi_j^*(\bold{r}_j)\boldsymbol{\lambda}_{j,\alpha}\hat{\bold{r}}_j %\chi_j(r_j)d\bold{r}_j\cdot \boldsymbol{\lambda}_{i,\alpha}\hat{\bold{r}}_i\bigg]\chi_i(\bold{r}_i)&=&\epsilon_i %\chi_i(r_i),\nonumber
%\end{eqnarray}
 Orthogonality conditions $\bra{\chi_n}\ket{\chi_m}=\delta_{nm}$ are automatically obeyed, i.e., our orbitals are automatically canonical, since we assume non overlapping orbitals. Eventually, we are ready to perform an ab-initio molecular dynamics simulation on the exactly dressed ground-state potential energy surface in classical canonical equilibrium by time-propagation of the standard Langevin equations of motion as given in Eqs. (6) and (7) of the letter. For the force calculation, we use that the Hellmann-Feynman theorem applies for the  variational cavity Hartree eigenvalue problem, provided that the local eigenvalue problem of the $n$-th molcecule can be solved with a variational method (as we do for the Shin-Metiu molecule by exact diagonalization below).

\section{Cavity ab-initio molecular dynamics for random oriented Shin-Metiu molecules }

To mimic the impact of rotational molecular disorder on collective vibrational strong couplig, we simulated rotational molecular motion at finite temperature by an overdamped rotational Langevin equation, i.b. by rotational Brownian motion.
This can efficiently be done for 1D Shin-Metiu molecules by introducing a time-dependent coupling constant
\begin{eqnarray}
    \lambda_{\alpha,n}(t) =\boldsymbol{\lambda}_\alpha \cdot  \bold{n}_{n}(t),\label{eq:lambdat}
\end{eqnarray}
for each individual molecule $n$. We have assumed a cavity polarization along the $z$-axis, i.e., $\boldsymbol{\lambda}_\alpha=\lambda_\alpha \bold{e}_z$ and that the 1D molecule should be aligned in 3D along the $\bold{n}_n(t)$-direction.
Notice that there is no feedback from the molecular and/or cavity state onto the normalized orientation $\bold{n}_n(t)$. In other words, we assume in the following that the rotational motion is solely determined by the  rotational Debey relaxation time $\tau_R=2\pi/ \omega_R$, which we will choose considerably smaller than the timescale of the nuclear vibrations (off-resonant with respect to the tuning of the cavity). However, the slowly changing random molecular orientation, will certainly influence the collective strong coupling effects between the molecules and the cavity. To mimic random rotational motion by $\bold{n}_n(t)$ the rotational Langevin equation  in the overdamped limit can be used, i.e.,  as,\cite{mazo2008brownian,marino2016entropy}
\begin{eqnarray}
    \frac{d \vec{n}_n(t)}{dt}=\sqrt{\tau_R}\vec{S}_n(t)\times\vec{n}_n(t)
\end{eqnarray}
with Debye relaxation time $\tau_R=2 k_B T/\eta_R$ and unbiased, delta-correlated Gaussian noise sources $\vec{S}(t)$. The rotational friction coefficient is labeled by $\eta_R$. This first order stochastic PDE is propagated numerically using forward Euler method in a similar spirit to Ref. \citenum{volpe2013simulation}, for which we yield
\begin{eqnarray}
   \vec{n}_{n,t+\Delta t}=\frac{\vec{n}_{n,t}}{|\vec{n}_{n,t}|}+\sqrt{\Delta t \tau_R}\vec{S}_{n,t}\times\frac{\vec{n}_{n,t}}{|\vec{n}_{n,t}|}.\label{eq:rotdif_num}
\end{eqnarray}
and exact normalisation is imposed at every time-step. Eqs. (\ref{eq:lambdat}) and (\ref{eq:rotdif_num}) then allow the simple simulation of vibrational strong coupling for a randomly oriented molecular ensemble within our self-consistent cBOA approach.  

\section{Simulation setup}

\subsection{Shin-Metiu molecular dynamics simulation}

The Hamiltonian operator $\hat{H}_{\rm SM}$ of the $n$-th Shin-Metiu molecule is given by,\cite{shin1995,Albareda2014}
\begin{eqnarray}
\hat{H}_{n}&=&\frac{\hat{P}^2}{2 M}+\frac{\hat{p}^2}{2}+\frac{1}{|L/2-\hat{R}|}+\frac{1}{|L/2+\hat{R}|}\nonumber\\
&&-\frac{\erf({|\hat{R}-\hat{r}|/R_f})}{|\hat{R}-\hat{r}|}-\frac{\erf({|\hat{r}-L/2|/R_r})}{|\hat{r}-L/2|}-\frac{\erf({|\hat{r}+L/2|/R_l})}{|\hat{r}+L/2|}.
\end{eqnarray}
We have used atomic units throughout our calculations with $R_f=R_l=R_r=1.511$ and $L=9.45$. We chose the proton mass $M=1836$ for the moving nuclei with positive unit charge and $m=1$ for the electron with negative unit charge. For the 1D electron, we chose a converged grid basis set representation  with 41 equidistant grid points and a grid spacing of $0.8$. Notice that having a large enough basis set for the electrons is pivotal to resolve local polarization effects.~\cite{yang2021quantum} The cavity Hartree equations were minimized self-consistently until converging the total electronic ensemble energy up to $\Delta E<1\times 10^{-7}$.  The classical Langevin equations of motion were propagated numerically using the scheme of Ref. \citenum{bussi2007accurate} with a time step $\delta t=50$. Trajectories were simulated over $2000$ time-steps to evaluate the local polarization effects and for $50000$ time-steps for the spectra calculations. Nuclei were initialized randomly distributed in the vicinity of the ground-state. Thermostating parameters were set to $k_B T=0.5\times 10^{-3}$ with low friction coefficient $\gamma=0.3\times 10^{-5}$ (underdamped regime). In case of randomly oriented molecules, the Debye relaxation time was set to $\tau_R=0.5\times 10^{-5}$.
Notice that the parametrization of the Shin-Metiu molecule and the temperature was chosen such that non-adiabatic coupling effects should not play a role, i.e., the groundstate cBOA approximation is valid. Furthermore, the temperature was chosen small enough that the thermal broadening does not interfere with our spectral interpretation. In addition, no chemical reaction occurs at such low temperatures  on the chosen time-scale (no proton transfer between the two energy minima of the Shin-Metiu molecule). The impact of local polarization effects on chemical reaction rates will be the focus of future work instead.

The vibrational absorption spectra was calculated using the power spectra method in Ref.\citenum{berens1981} with a Blackman filter window\cite{blackman1958measurement} averaged over $33$ overlapping trajectory windows containing $4096$ time steps, each of them shifted by $1/3$ of the window size. While for the global absorption spectrum the total dipole (electronic + nuclear contribution) were post-processed accordingly (doted lines), we did the same for each individual molecular dipole in case of the local spectra calculation instead. Afterwards, the summation over all local spectra was taken yielding the bold lines in the spectral figures.   

\begin{figure}
          \begin{subfigure}[b]{0.45\textwidth}
         \centering
         \includegraphics[width=\textwidth]{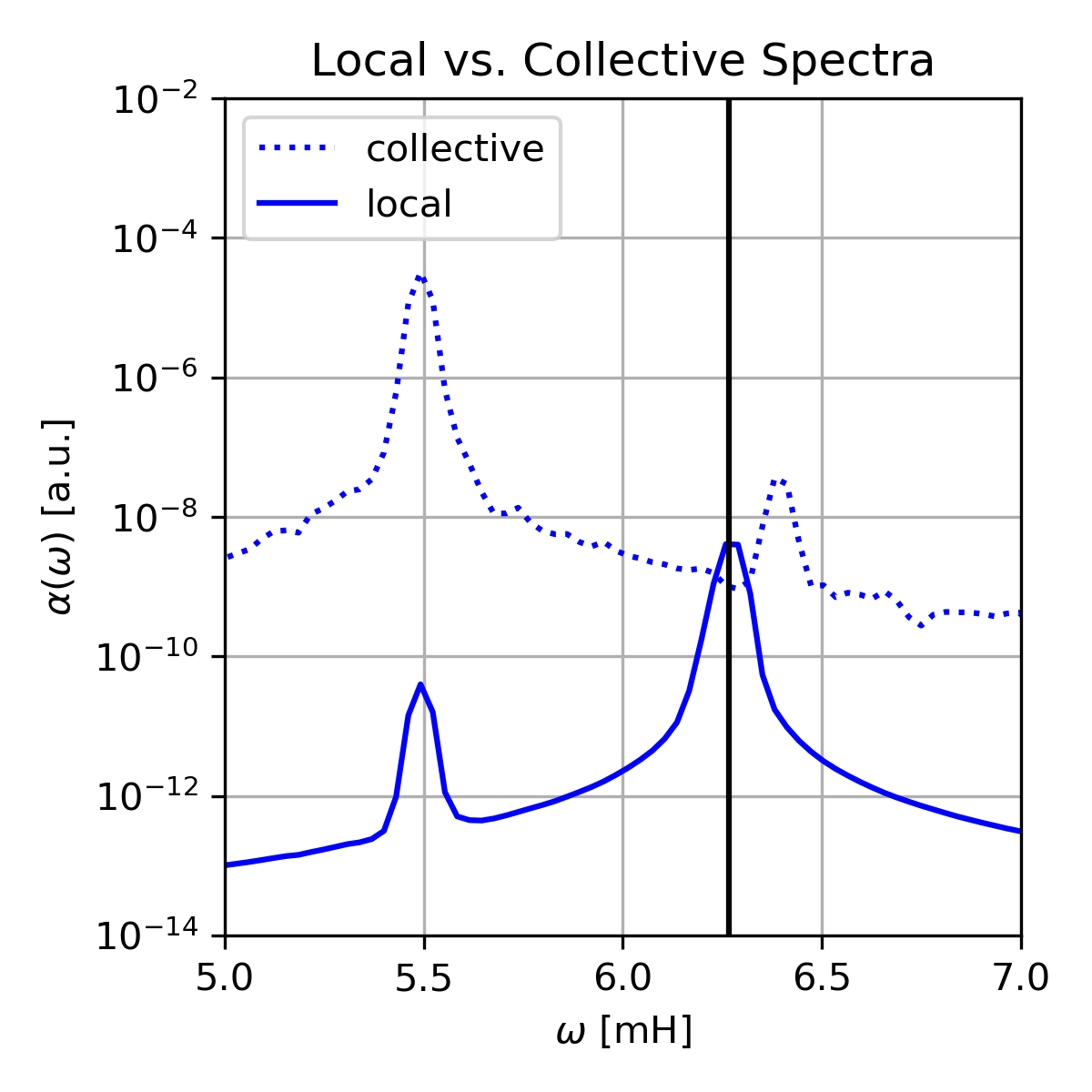}
         \caption{}
         \label{fig:y equals x}
     \end{subfigure}
      \begin{subfigure}[b]{0.45\textwidth}
         \centering
         \includegraphics[width=\textwidth]{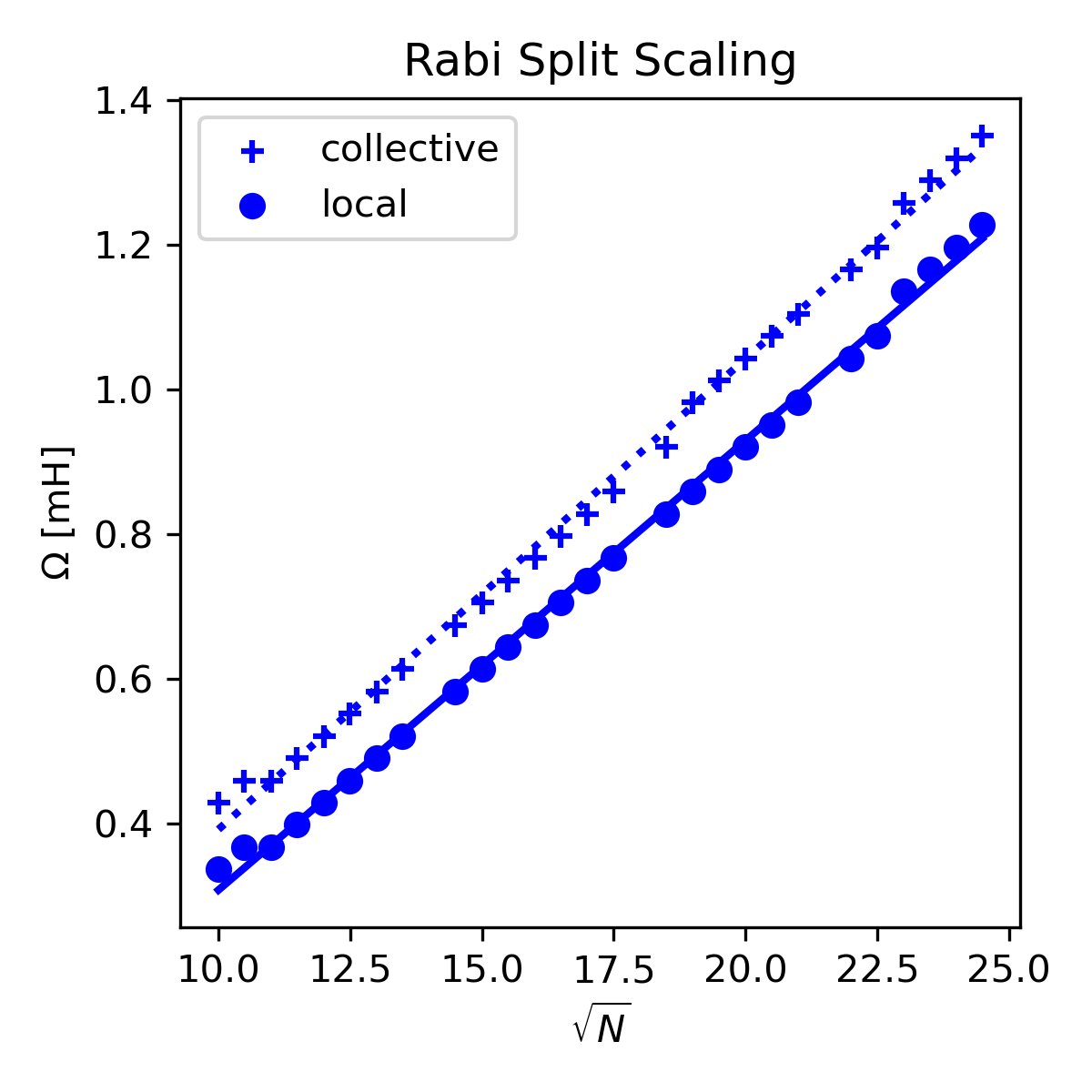}
         \caption{}
         \label{fig:y equals x}
     \end{subfigure}
        \caption{(a) Vibrational absorption spectra $\alpha(\omega)$ for 900 aligned Shin-Metiu molecules under collective vibrational strong coupling similar to the random oriented molecules shown in the letter. The cavity is again tuned on the first vibrational mode of the uncoupled molecule at $\omega_\alpha=6.27$ [mH] (black vertical line). However, the coupling is reduce to $\lambda_\alpha=0.00425$ in order to achieve a Rabi splitting of similar magnitude. We find qualitative identical results for the aligned molecules as for the random oriented setup, i.e. a de-tuning of the cavity to lower frequencies in combination with a locally populated lower polaritonic state. \\
        (b) The collective $\sqrt{N}$-scaling law of the Rabi splitting remains preserved, when solving the self-consistently the coupled electronic problem for the molecules under vibrational strong coupling. Similarly, the observed local splitting follows the same $\sqrt{N}$-scaling law, where we defined the splitting between (local) lower polariton and the dark states located at $\omega_\alpha$. This suggests that local strong coupling can be increased by  collectively enhancement of the coupling. Small discrete patterns in the data emerge from finite spectral resolution of the discrete Fourier transformation applied to dipole trajectories of finite length.}
        \label{fig:delta_re}
\end{figure}

\begin{figure}
     %\begin{subfigure}[b]{0.45\textwidth}
         \centering
         \includegraphics[width=0.45\textwidth]{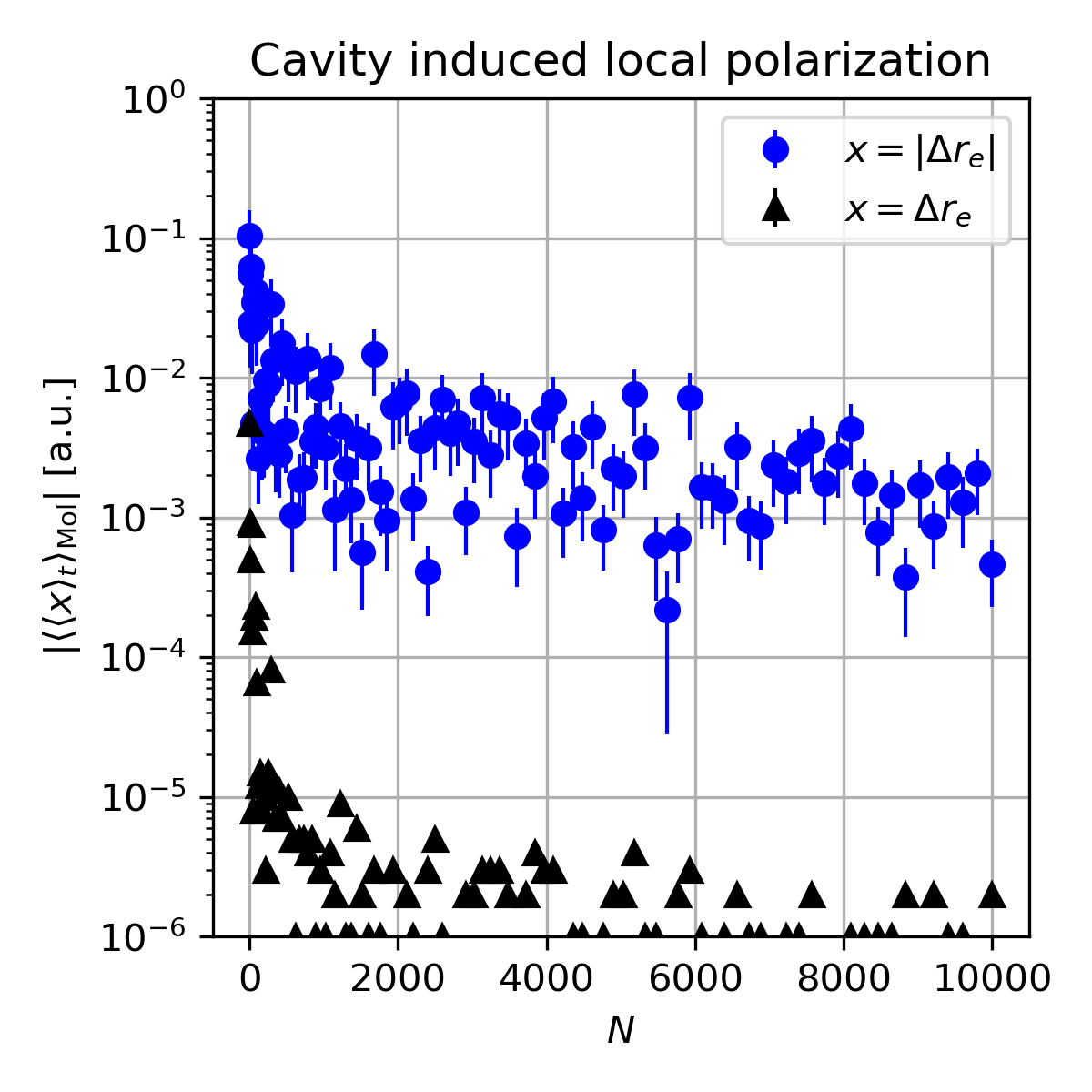}
         \caption{Statistical evaluation of the  electronic polarization $\Delta r_n(t)=\langle \hat{r}_n\rangle_{\lambda}-\langle \hat{r}_n\rangle_{\lambda=0}$ of the $n$-th molecule with respect to the bare molecule Shin-Metiu molecule in canonical equilibrium at $k_B T=0.5$ [mH] for aligned molecules. The Rabi-splitting was kept constant when increasing the number of molecules by choosing a re-scaled $\lambda_\alpha(N)=0.256/\sqrt{2 N}$. By monitoring $|\Delta r_n|$ (blue dots), we observe qualitatively similar results comparing with the random oriented molecules. The simulations suggest a non-zero saturation of the cavity induced local polarizations in the large $N$ limit, where the standard deviations with respect to different molecules are displayed by vertical blue lines. At the same time, the total polarization of the ensemble, which is related to $\Delta r_n$ (black triangle), quickly approaches zero (exactly zero when averaging over longer time-scales), since the cavity cannot induce a non-zero ensemble polarization in thermal equilibrium.    
         Consequently, our simulations suggest that cavity induced local strong coupling effects persist in the thermodynamic limit $(N\rightarrow\infty)$ of a molecular ensemble under collective vibrational strong coupling. In other words, the self-consistent treatment is decisive to describe ground-state polaritonic chemistry accurately for collectively coupled molecular ensembles. }
         \label{fig:dre}
     %\end{subfigure}
%          \begin{subfigure}[b]{0.45\textwidth}
%         \centering
%         \includegraphics[width=\textwidth]{Delta_re_t_mol_N_temperature_00005.png}
%         \caption{}
%         \label{fig:y equals x}
%     \end{subfigure}
\end{figure}

Supplementary simulation results for aligned Shin-Metiu molecules (spectra, Rabi-split scaling and local polarizations) are shown in Figs. \ref{fig:delta_re} and \ref{fig:dre}.

\subsection{Ensembles of HF molecules and Ne atoms}

The cBOA Hartree-Fock method and corresponding analytic nuclear gradients were implemented in the Psi4NumPy environment~\cite{Smith2018-tu,schnappinger2023cavitybornoppenheimer}, which is an extension of the PSI4~\cite{Smith2020-kq} electronic structure package. All calculations were performed using the aug-cc-pVDZ basis set~\cite{Kendall1992-wu} and the geometry of the isolated single HF molecule is optimized at the Hartree-Fock level of theory. Note that we have not reoptimized the geometries of the molecular systems in the cavity; as such, our calculations do not account for any geometric relaxation effects stemming from the presence of the cavity. In all cavity Hartree-Fock calculations performed in this work, we consider a single-mode and non-lossy cavity. The fundamental cavity frequency $\omega_{\alpha}$  is tuned to the first vibrational mode of the uncoupled HF at $20.35$ [mH].

For the molecular ensembles studied, the optimized structure of a single HF molecule is replicated $N$ times. All these replicas are separated by $800$~\AA~and placed inside a cavity. All individual molecular dipole moments are aligned with the cavity polarization axes, and the zero transversal electric field condition is satisfied for the whole ensemble. The same distance of $800$~\AA~is used for the small ensembles of neon atoms. The vibrational spectra for the cavity-coupled ensembles are calculated in the harmonic approximation using numerical second derivatives of the cavity Hartree-Fock energy. All calculations were performed in a reproducible environment using the Nix package manager (nixpkgs, 22.11, commit 594ef126) in combination with NixOS-QChem (commit f5dad404).\cite{nix}

\begin{figure}
         \centering
         \includegraphics[width=0.6\textwidth]{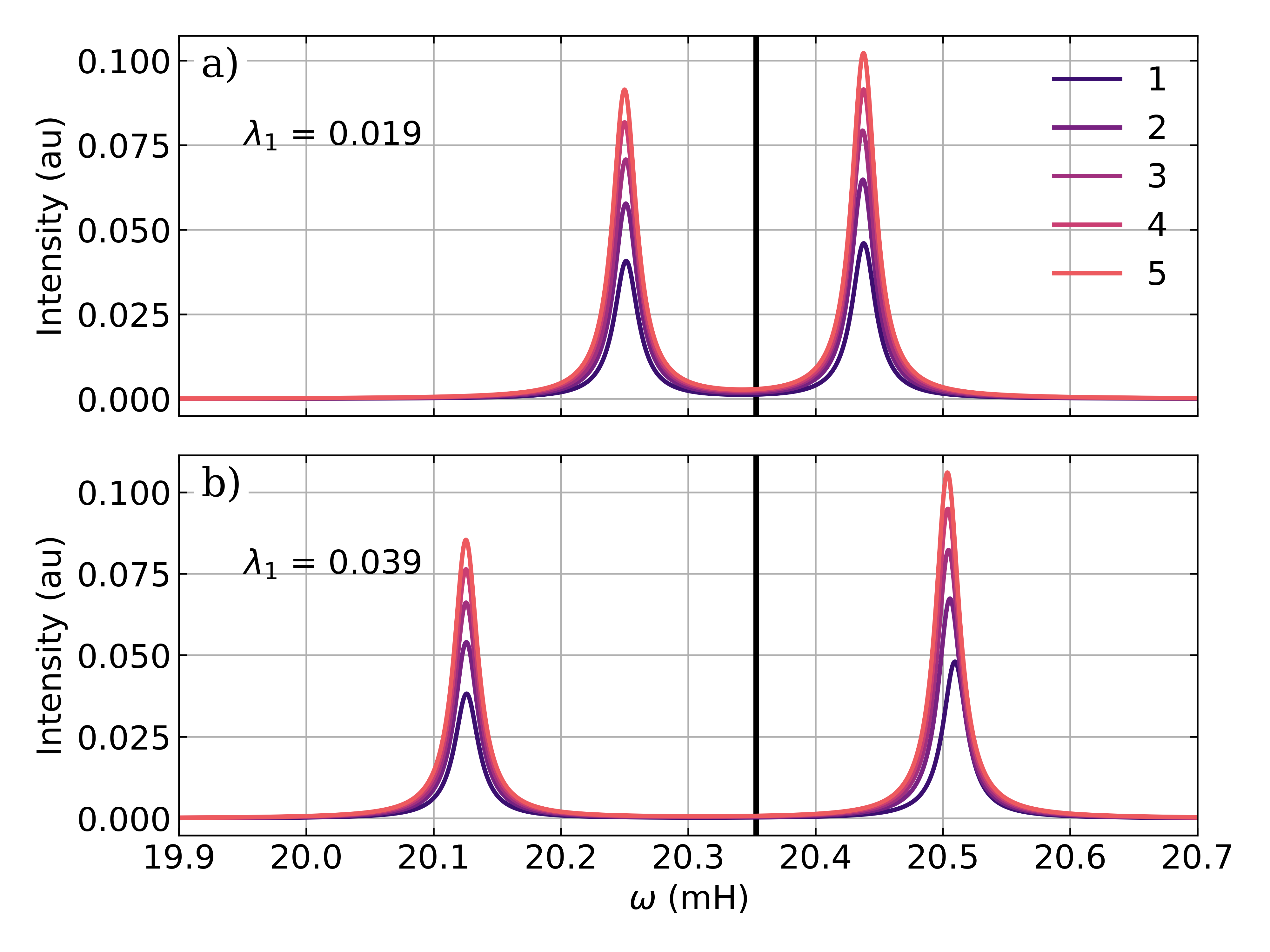}
        \caption{Vibrational absorption spectra in the harmonic approximation  for a few perfectly parallel aligned HF molecules under collective vibrational strong coupling. The number of molecules is color coded. The cavity is tuned to the first vibrational mode of the uncoupled HF molecule at $\omega_\alpha=20.35$ [mH] (black vertical line) with unscaled coupling strength of a) $\lambda_1=0.019$ and b) $\lambda_1=0.039$. }
        \label{fig:hf_spectra}
\end{figure}

Supplementary simulation results for the vibrational absorption spectra of the molecular ensembles are shown in Fig. \ref{fig:hf_spectra}.

\begin{figure}
         \centering
         \includegraphics[width=0.6\textwidth]{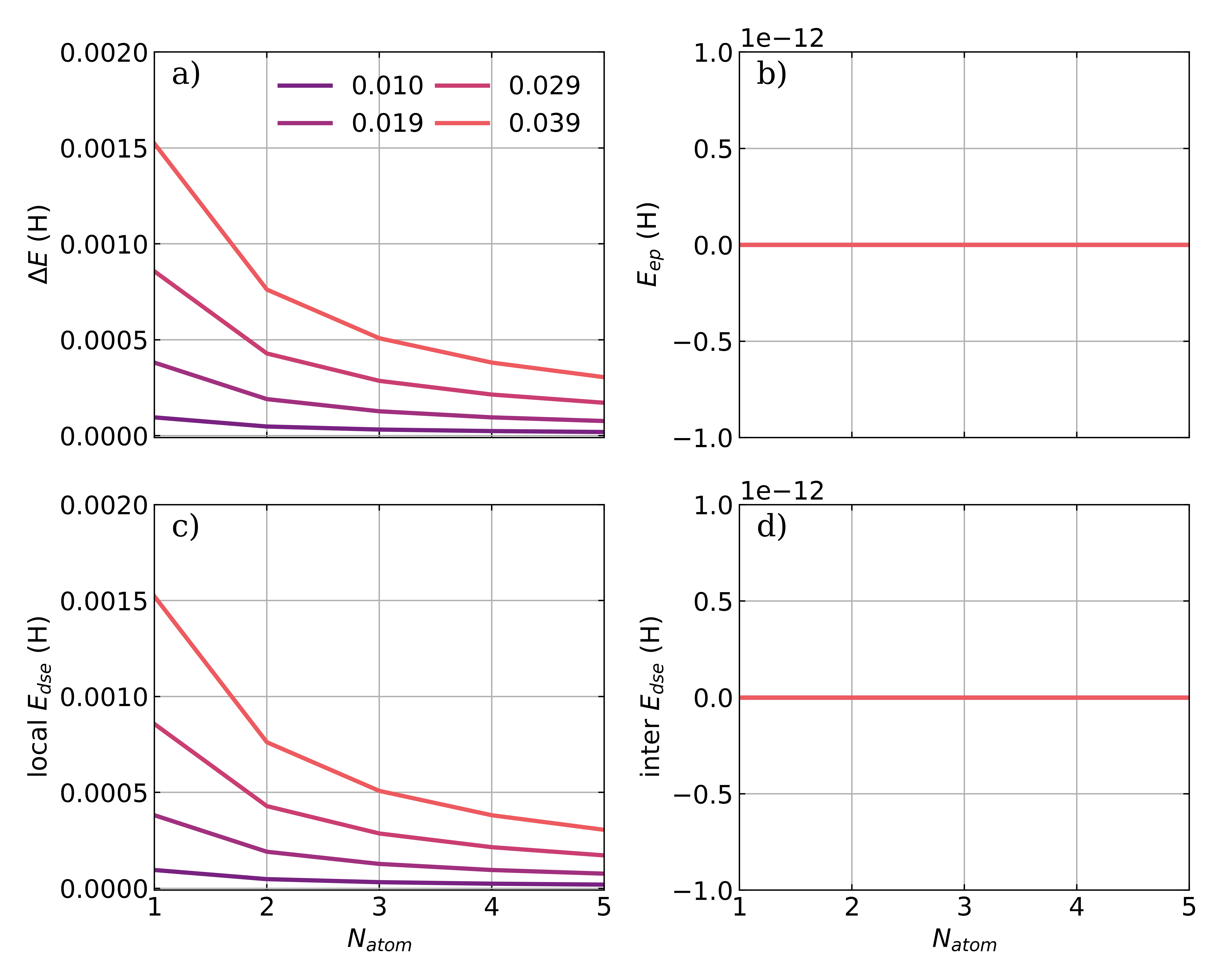}
        \caption{Influence of the cavity interaction on an individual neon atom in ensembles of different size. a) The total energy change with respect to the cavity free situation, b) the local contribution of the cavity-induced photon displacement  $E_{ep}$, c) the local cavity induced polarization contribution $E_{dse}$ and d) the dipole-dipole interaction term $V_{dd}$ as a function of $N_{atom}$.  A cavity frequency $\omega_{\alpha}$ of $20.35$ [mH] is used. The value of the unscaled coupling strength $\lambda_1$ is increased from 0.010 to 0.039 (color coded).}
        \label{fig:atomic_local}
\end{figure}

Local energy modifications for the ensembles of neon atoms are shown in Fig.~\ref{fig:atomic_local}. Since atoms do not have a permanent dipole moment, the energy contribution of the cavity-induced photon displacement $\hat{\boldsymbol{D}}$ and the dipole-dipole interaction term $V_{dd}$ are exactly zero, see Fig.~\ref{fig:atomic_local}~b) and d). Only the local cavity-induced polarization, see (Fig.~\ref{fig:atomic_local}~c), give rise to  a non-zero contribution to the locale energy (see Fig.~\ref{fig:atomic_local}~a)) for the investigated ensemble sizes. However, this contribution decays with $\frac{1}{N}$ and is negligible in the large-$N$ limit. 

\bibliography{si}